\documentclass[aps,superscriptaddress,11pt]{revtex4}
\usepackage{graphicx}
\usepackage{amsmath,amsthm, amssymb,latexsym, amsfonts}
\usepackage{color}
\usepackage{longtable}
\newcommand{\nobel}{\textcolor{red}{*}}
\newcommand{\paperintext}[1]{\textit{``#1''}}

\begin{document}

\title{The citation wake of publications detects Nobel laureates' papers}
\author{David F. Klosik}
\author{Stefan Bornholdt}
\email[email:]{bornholdt@itp.uni-bremen.de}
\affiliation{Institut f\"ur Theoretische Physik, Universit\"at Bremen, 
Hochschulring 18, 28359 Bremen}

\maketitle

{\bf
For several decades, a leading paradigm of how to quantitatively 
assess scientific research has been the analysis of the aggregated 
citation information in a set of scientific publications. 
Although the representation of this information as a citation 
network has already been coined in the 1960s \cite{deSollaPrice1965}, 
it needed the systematic indexing of scientific literature to allow 
for impact metrics that actually made use of this network as a whole  
\cite{chen_etal_2006,walker_etal_2007,Radicchi_et_al_2009}, 
improving on the then prevailing metrics that were almost 
exclusively based on the number of direct citations 
\cite{Garfield1992,Hirsch2005}. 
However, besides focusing on the assignment of credit, 
the paper citation network can also be studied in terms of 
the proliferation of scientific ideas. 
Here we introduce a simple measure based on the shortest-paths 
in the paper's in-component or, simply speaking, on the shape 
and size of the \textit{wake} of a paper within the citation network. 
Applied to a citation network containing \textit{Physical Review} 
publications from more than a century, our approach is able to 
detect seminal articles which have introduced concepts of obvious 
importance to the further development of physics. 
We observe a large fraction of papers co-authored by Nobel 
Prize laureates in physics among the top-ranked publications.
}
\bigskip

To a certain degree science is a system of proliferation of ideas, 
since even the most innovative concepts emerge as answers to open 
problems and are based (at least to some extent) on previously 
established methods: they explain seemingly incompatible data 
within a new theoretic framework, or they make data accessible in the
first place by rendering new measurements or calculations feasible. 
Fortunately, the associated book-keeping is done by scientists 
themselves due to the practice of citing the work they were 
influenced by in the reference list of their own publications. 
The aggregated citation information in a set of scientific 
publications therefore allows for deep insight in the structure 
and progress of science. As early as the 1960s it has first
been depicted as a paper citation network, in which papers are 
represented by nodes and citations by directed links between 
the associated nodes of a citing paper and the corresponding 
cited paper \cite{deSollaPrice1965,deSollaPrice1976}.

However, during the past decades, citation analysis has rather 
focused on the quantitative assessment of scientific research, 
interpreting citations as indicators of impact or assignment of 
credit \cite{Petersen_et_al_2010}. To this end, numerous 
quantitative measures have been developed \cite{Bollen_etal_2009}, 
many of which are designed to rank scholars
(e.g. the h-index \cite{Hirsch2005}) or journals 
(e.g. the Thomson Scientific Journal Impact Factor). 
Until recently these measures were almost exclusively based on
counting the number of a paper's direct citations (i.e.\ the 
respective node's in-degree), which is, indeed, known to have 
several shortcomings that have been discussed in the literature 
\cite{Hamilton1991,MacRobertsMacRoberts1989,RadicchiCastellano2012,RadicchiCastellano2011}. 
Motivated by the growing accessibility of citation databases, 
over the last decade some of these drawbacks have been addressed 
by a number of more elaborate impact measures that do not restrict 
themselves to mere direct citations (e.g. CiteRank \cite{walker_etal_2007}, 
SARA \cite{Radicchi_et_al_2009}, Eigenfactor \cite{West_etal_2010}), 
but consider contributions to a paper's impact other than receiving 
the community's attention \cite{Wu_Huberman_2007}. 
In particular, Chen et al. have shown that Google's PageRank algorithm 
\cite{BrinPage1998,GhoshalBarabasi2011}, that can be thought of as a 
diffusion process in which credit is equally distributed among 
the out-neighbours of a node, is able to find groundbreaking old 
papers ('scientific gems'), that are rather moderately cited directly 
\cite{chen_etal_2006}. 

Focusing on the picture of idea propagation rather than of credit diffusion, 
we here introduce an approach to ranking scientific publications in a 
paper citation network that is based on an intuitive interpretation 
of the set of all publications that have cited a paper in question 
directly or indirectly (i.e.\ the paper's in-component).

Let $G$ be a paper citation network that is composed of the aggregated 
information about $m$ citations between $n$ papers, and let $i$ be 
a paper that has been published at a point in time $t_i$. 
Clearly, paper $i$ can only be cited by papers published at a later 
date $t_j > t_i$, which again can receive citations only from papers 
published later than $t_j$ and so on, making $G$ a directed acyclic 
graph (DAG). The wake-citation-score, $w_i$, of a publication $i$ is 
based on assigning all papers in $i$'s in-component, which we here 
call its \textit{wake} $\mathbb{W}_i$, to \textit{neighbourhood layers} 
$\mathbb{L}_{i,k}$ according to the length $k+1$ of the shortest-path(s) 
to $i$ with $k=0$ corresponding to the direct neighbours.
Note that the resulting layers are disjoint and papers appear only in the closest one. 
The score $w_i$ is then computed as a weighted sum over the 
cardinalities of the layers with the addends damped with a dilution factor 
depending on a parameter $\alpha \in [0,1]$, here chosen to be $\alpha^k$. 
While for $\alpha=0$ the computation is restricted to the zeroth layer, 
i.e.\ the direct citations, for $\alpha=1$ the whole wake is considered 
without any damping.

\begin{figure}[htb]
\includegraphics[width=.97\columnwidth]{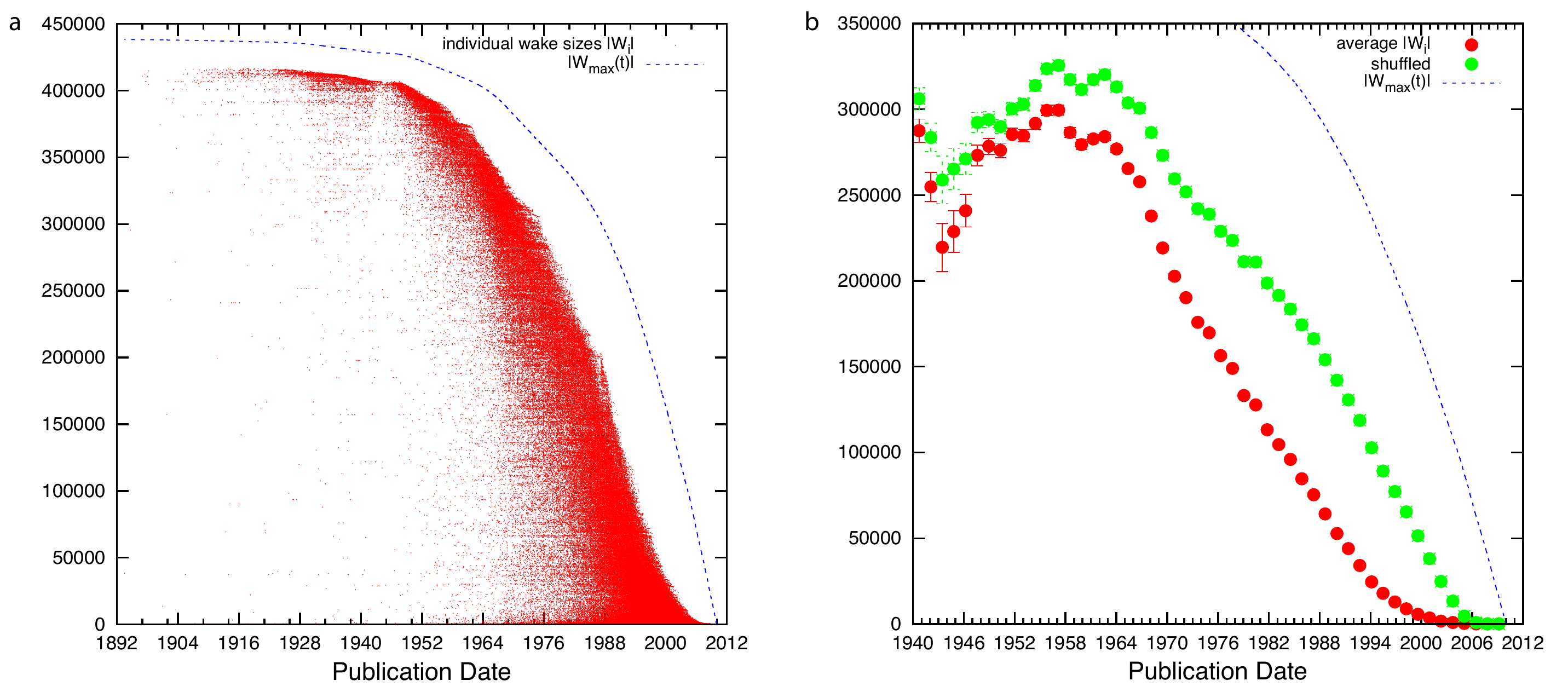}
\caption{ \label{fig:volleNachlaeufe}
{\bf The papers' in-components. }
{\bf a,} The in-component sizes, $|\mathbb{W}_i|$, of all 
papers in the dataset are shown over the publication date. 
The dashed line depicts the size of the maximal in-component 
at a given publication date, $|\mathbb{W}_{\text{max}}(t)|$. 
Early publications rarely show intermediate $|\mathbb{W}_i|$ 
but are either rather poorly cited or have an in-component 
of $90$ percent of the maximum wake. The resulting 'ridge' 
formed by the data indicates cross-references between 
scientific subfields. 
For more recently published papers a subfield-structure is 
implied by the much smaller fraction of the maximum 
$|\mathbb{W}_i|$ at a given time and the corresponding
$|\mathbb{W}_\text{max}(t_i)|$. 
In {\bf b} the data is averaged over time-windows of 
$500$ days (red) and compared to wake-sizes obtained from 
a shuffled version (green) of the citation network,
the shuffling being applied via a switching method 
that conserves the inherent time-arrow and the degree 
sequence of the network \cite{MaslovSneppen_2002}. 
The link-shuffling leads to larger average in-component 
sizes for younger papers: the subfield-structure is not 
preserved in the shuffled network.
}
\end{figure}

Although Fig.~\ref{fig:volleNachlaeufe} shows that the total wake sizes 
yield information about the overall structure of the citation network 
(e.g. they indicate the presence of cross-references between different 
scientific subfields \cite{ChenRedner2010}), it is the case of 
$\alpha < 1$ that reflects our notion of idea dilution in the network: 
The only explicit source of dilution in the wake-citation-score is 
given by the length of the shortest-path(s) between two papers, 
i.e.\ the minimal number of processing steps of an idea. 
In particular, the contribution to the score of a cited paper $i$ 
is not normalized by the out-degree of the citing paper $j$ 
(as it is the case in the PageRank algorithm), since this would 
correspond to reducing the influence of $i$ on $j$ due to the mere 
presence of other sources of ideas that $j$ has used. 

Eventually, a detrending is applied in order to address the fact 
that in a finite citation network there are the more possible citing 
papers the earlier a paper $i$ has been published. 
We normalize the wake-citation-score by the size of the maximum 
possible in-component of a paper published at the time of $i$'s 
publication, $\mathbb{W}_\text{max}(t_i)$, i.e.\ the number of 
all papers that are published after $t_i$ and arrive at the expression
\begin{equation}
  w_i = \frac{1}{ \left| \mathbb{W}_{\text{max}}(t_i) \right| } 
  \sum_k \alpha^k  \left| \mathbb{L}_{i,k} \right|.
  \label{eq:wakescore}
\end{equation}
Note that this is a sum over disjoint layers of nodes emphasizing 
that the wake-citation-score counts weighted nodes and not weighted 
paths as centrality measures like PageRank do: every paper 
$j\in \mathbb{W}_i$ is evaluated exactly once in the computation 
of paper $i$'s score, while there is a contribution of $j$ to the
PageRank score of $i$ for every directed path from $j$ to $i$.

Since the assigning of disjoint neighbourhood layers corresponds 
to the construction of a shortest-paths-tree spanning the subgraph 
$G_i$ which contains all nodes in $\mathbb{W}_i$ and the associated links, 
computing the wake-citation-scores for all nodes can be done by 
solving the all-pairs shortest-paths problem (APSP) and is thus 
of order $\mathcal{O}(mn+n^2)$ when a Breadth-First-Search is applied.

As mentioned above, the direct citation count is biased in several ways. 
In particular, it will strongly depend on the citation habits in the
respective field and the size of the latter. Furthermore, within the 
period covered by the dataset used here (July 1893 - December 2009) 
there has been a remarkable growth of the scientific community as a whole. 
A paper published in the 1920s can potentially be cited by a much 
larger number of papers than a publication from the 1990s, 
but as the propability of being cited decays in time \cite{redner_2005}, 
one should also consider the number of papers published in the 
adjacent years when receiving citations is most likely: 
While there are not more than $1663$ papers published in the 
whole 1920s in the dataset, from 1992 on in each single year 
more than 10,000 papers are published. 
Also considering the fact that some seminal results have become 
textbook-knowledge, so that the original papers are no longer cited, 
the direct citation count might underestimate the impact of 
groundbreaking old publications. The wake-citation-score presented 
here is designed to address these issues. 

\begin{figure}[htb]
\includegraphics[width=.97\columnwidth]{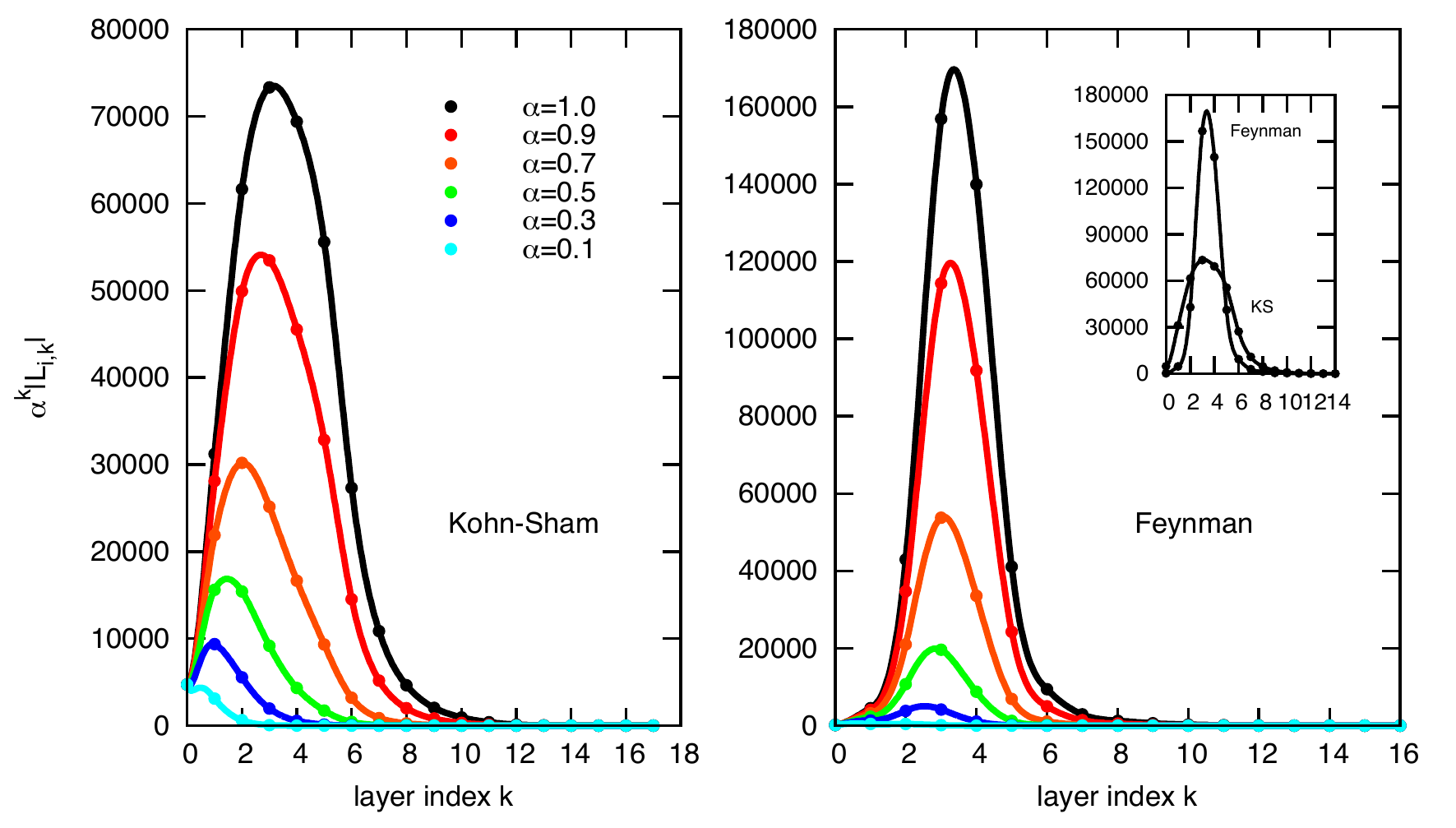} 
\caption{ 
{\bf Neighbourhood layers of the citation wakes of two articles.}
The dilution concept in the computation of the wake-citation-score 
is illustrated by the cardinalities of the neighbourhood layers of 
two individual articles \cite{KohnSham1965,Feynman1949}. 
In black the mere numbers of neighbours in the k-th layer are given 
($\alpha=1$), while other colours correspond to an actual dilution 
with $\alpha<1$. Note that the number of direct citations 
(at layer index $k=0$) is large for the \textit{Kohn-Sham}- 
but rather small for the \textit{Feynman}-paper.
The wake-citation-score for a specific $\alpha$ can be visually 
approximated by the area under the associated curve.
}
\label{fig:vergleichsbild}
\end{figure}

The three top-cited publications in the dataset are 
\paperintext{Self-Consistent Equations Including Exchange and 
Correlation Effects} by W.\ Kohn and L.\ Sham \cite{KohnSham1965}, 
\paperintext{Inhomogeneous Electron Gas} by P.\ Hohenberg and W.\ Kohn 
\cite{HohenbergKohn1964}, and \paperintext{Self-interaction correction to
density-functional approximations for many-electron systems} 
by J.\ Perdew and A.\ Zunger \cite{PerdewZunger1981} (cited $4728$, 
$3682$, $3169$ times, respectively). Fig.~\ref{fig:vergleichsbild} 
shows the individual wake of the Kohn-Sham paper. 
They all are seminal papers in the density-functional theory, 
the former two, indeed, prepared the ground for the method by 
introducing the Kohn-Sham equations and the Hohenberg-Kohn theorem. 
The number of direct citations seems to adequately represent their importance. 
But note the publication shown in the right frame of 
Fig.~\ref{fig:vergleichsbild}: In his 1949-paper 
\paperintext{Space-Time Approach to Quantum Electrodynamics}
R.\ Feynman introduced what became famous as the Feynman diagram 
\cite{Feynman1949}. Despite the undisputable fruitfulness of this 
concept the associated paper is rather moderately cited ($206$ times) 
and, considering direct citations only, is ranked $\#1031$. 
While the citation count clearly underestimates its influence,
applying the wake-citation-score yields one of the top-ranks 
($\#6$ for $\alpha=0.9$, see below).

As a matter of course, there is no objective measure of importance 
or impact that the results of a bibliometric ranking scheme could 
be compared with. However, in order to argue that the 
wake-citation-score yields plausible results, in Table 
\ref{tab:alpha09} we show the ten top-ranked papers according to 
their $w_i$-score with the parameter $\alpha$ chosen to be $0.9$. 
Besides the aforementioned seminal publication by R.\ Feynman 
(ranked $\#6$) the list is mainly composed of further groundbreaking 
papers whose high score is obviously plausible. 
Note for example the publication \paperintext{The Radiation Theories 
of Tomonaga, Schwinger, and Feynman} in which F.\ Dyson shows the 
equivalence of Feynman's diagrammatic approach to quantum 
electrodynamics introduced in \cite{Feynman1949} with the formalism 
used by J.\ Schwinger and S.-I.\ Tomonaga. 
The concept of the Wigner-Seitz-cell, which has become common knowledge 
decades ago, was presented in the paper \paperintext{On the Constitution of
Metallic Sodium} by E. Wigner and S. Seitz in 1933, which is ranked $\#7$ here. 
With the determinant representation of a basis of the anti-symmetrized 
fermionic many-body-wavefunction J. Slater had introduced an even 
more fundamental concept in 1929. His publication \paperintext{The Theory 
of Complex Spectra} \cite{Slater1929} is ranked $\#10$ here.
The remaining articles also seem properly placed within the top-ranked 
publications.
However, except the highest ranked paper by Bardeen, Cooper and Schrieffer 
the ten top-ranked papers are far from being amongst the ten top-cited 
papers, in fact most of them are rather cited moderately and would not 
stand out remarkably in a direct citation-based ranking. 
The value given in the second column of Table~\ref{tab:alpha09} 
illustrates the wake-citation-score's intriguing feature of finding 
these articles: All shown publications show high ratios between the 
direct citation-rank and the rank assigned according to the 
wake-citation-score (Dyson's paper and Slater's \paperintext{The 
Theory of Complex Spectra} are especially remarkable examples).
Indeed, amongst the top-100 papers $86$ show a ratio exceeding a 
value of $10$. 

\begin{figure}[ht]
 \includegraphics[width=0.9\columnwidth]{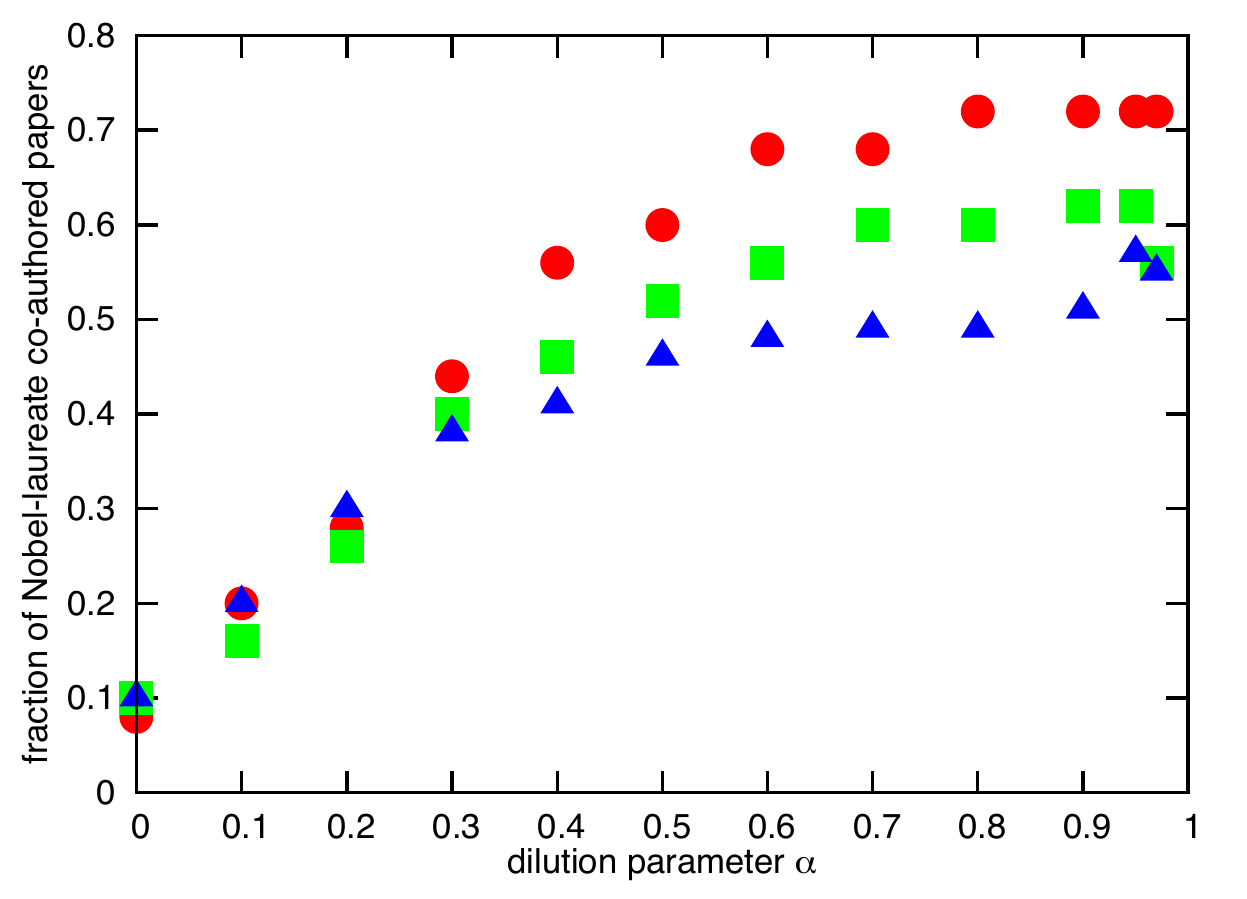} 
 \caption{\label{fig:alphascan}
 {\bf Fraction of Nobel-laureate co-authored papers.}
 Fraction of papers with a co-author who has been awarded 
 with a Nobel Prize in physics among the top-$25$ (red circles), 
 top-$50$ (green squares) and top-$100$ (blue triangles) 
 ranked papers according to the wake-citation-score for 
 different dilution parameter values $\alpha$.
 }
\end{figure}

As a second plausibility check of the outcome provided by the 
wake-citation-score we checked the top-ranked papers for the 
co-authorship of Nobel Prize laureates \cite{Mazloumian_et_al_2011} 
in physics, which we consider to serve as a first-approximation 
proxy for both scientific quality and - although to a smaller extent, 
as there can be multiple papers co-authored by the same laureate 
in the list - diversity concerning physical subfields. 
In Table~\ref{tab:alpha09} the Nobel Prize laureates are 
labelled with a red star.
Fig.~\ref{fig:alphascan} illustrates the dependence on the 
dilution parameter $\alpha$; since for a dilution parameter 
$\alpha=1$ the wake-citation-score forfeits its ability to 
distinguish between different shortest-path lengths, i.e.\ 
different numbers of processing steps of an idea, we only show 
$\alpha$ values from the interval $[0,1)$.
We found remarkable high fractions of papers co-authored by at 
least one physics Nobel Prize laureate among the $25$, $50$ and 
$100$ top-ranked publications: e.g.\ for $\alpha=0.9$ the list 
of the top-$25$ publications contains $18$ of such papers ($72\%$), 
$31$ of them are among the top-raked $50$, and still more than 
half of the top-100-ranked publications have been contributed 
to by a Nobel Prize laureate. This is in contrast to the ranking 
according to the direct citation count: not more than $4$, $10$ 
and $25$ laureate-co-authored papers are among the top-25, top-50
and top-100 cited papers, respectively. 
We also compared our results to the outcome of a ranking obtained 
by the application of PageRank to the network as described in \cite{chen_etal_2006} 
and found that while PageRank also has more Nobel laureate 
co-authored papers among the top-ranked publications than the 
direct citation score, the wake-citation-score yields higher values 
for a wide range of the dilution parameter $\alpha$ whose  
value is not critical and does not require tuning. 

In summary, our approach of decomposing the papers' in-components 
into neighbourhood layers succeeds in finding groundbreaking 
publications that are rather moderately cited as well as in yielding 
a more diverse ranking than the direct citation count 
(which is dominated by the density-functional theory), 
without any explicit consideration of the lengths of the papers' 
reference lists and without an \textit{a pirori} assignment of 
the papers to specific subfields (as it is the case with relative 
indicators \cite{RadicchiFortunatoCastellano2008}).

\textit{ }\newline
{\bf Methods}\newline
We apply the wake-citation-score to a paper citation network 
which is composed of papers published in the APS-journals between
July 1893 and December 2009 and the associate citation information 
(which can be requested at \url{https://publish.aps.org/datasets}). 
Note that only internal citations (between APS-publications) are provided. 
In order to exclude publications that presumably show non-typical 
citation patterns, we restrict the citation network nodes to 
standard publications (which are tagged \textit{article}, \textit{brief}, 
\textit{rapid} and \textit{letter} in the APS-dataset) and reject 
material tagged \textit{erratum}, \textit{comment}, \textit{reply}, 
\textit{miscellaneous}, \textit{editorial}, \textit{announcement}, 
\textit{essay}, \textit{publisher-note}, \textit{retraction}. 
The latter does not account for more than about $5$ percent of all 
publications in the dataset. In addition, we discard self-links 
and citations that do not point backward in time, as well as parallel 
edges, and obtain an acyclic directed graph with $438,296$ nodes 
and $4,560,230$ edges.

\textit{ }\newline
{\bf Acknowledgements}\newline
We thank the American Physical Society for allowing us to use 
their citation database of APS-journals. We further acknowledge 
support of the Deutsche Forschungsgemeinschaft under contract 
No. INST 144/242-1 FUGG.

\newpage

\begingroup
\squeezetable
\begin{longtable*}{| c | l | l l |} 
\hline
$\text{rank}(w_i)$ & $\frac{\text{rank}(c_i)}{\text{rank}(w_i)}$ & \multicolumn{2}{c |}{ Publication } \\
\hline
\hline
	$1$ & $10$ & Bardeen\nobel, Cooper\nobel, Schrieffer\nobel & Theory of Superconductivity 		\\
	$2$ & $707.5$ &F. Dyson & The Radiation Theories of Tomonaga, Schwinger, and Feynman \\
	$3$ & $30.7$ &E. Wigner\nobel & On the Interaction of Electrons in Metals  	\\
	$4$ & $90.6$ &M. Gell-Mann\nobel, K. Brueckner & Correlation Energy of an Electron Gas at High Density 	\\
	$5$ & $18$ & S. Chandrasekhar\nobel & Stochastic Problems in Physics and Astronomy 		\\
 	$6$ & $171.8$ & \textcolor{blue}{R. Feynman}\nobel & \textcolor{blue}{Space-Time Approach to Quantum Electrodynamics}\\
	$7$ & $166.9$ &E. Wigner\nobel, F. Seitz & On the Constitution of Metallic Sodium 		\\
	$8$ & $268.4$ &R. Feynman\nobel & The Theory of Positrons 		\\
	$9$ & $45.3$ &J. Slater & Atomic Shielding Constants 		\\
	$10$& $294.5$ & J. Slater & The Theory of Complex Spectra  \\ \hline
\caption{\label{tab:alpha09}
{\bf Top-10 ranked publications.} 
The ten top-ranked publications according to the wake-citation-score 
with dilution parameter $\alpha$ chosen to be $\alpha=0.9$. 
Nobel Prize laureates are labelled with a red asterisk (\nobel), 
the blue marked publication is also depicted in Fig.~\ref{fig:vergleichsbild}. 
The second column shows the fraction of the ranks assigned 
to the paper according to the number of direct citations and 
the wake-citation-score, respectively.} 
\end{longtable*}
\endgroup

\newpage

\end{document}